\documentclass{conm-p-l}

\newtheorem{theorem}{Theorem}[section]
\newtheorem{lemma}[theorem]{Lemma}

\newtheorem{prop}[theorem]{Proposition}

\theoremstyle{definition}
\newtheorem{definition}[theorem]{Definition}

\theoremstyle{remark}
\newtheorem{remark}[theorem]{Remark}

\numberwithin{equation}{section}

\newcommand{\supp}{\operatorname{supp}}

\begin{document}

\title{On the regularity of the density of electronic wavefunctions}

%    Information for first author
\author{S. Fournais}
%    Address of record for the research reported here
\address{This work was partially carried out at The Erwin
Schr\"odinger International
           Institute for Mathematical Physics,
           Boltzmanngasse 9,
           A-1090 Vienna,
           Austria}
%    Current address
\address{S. Fournais: Laboratoire de Math\'{e}matiques,
           Universit\'{e} Paris-Sud - B\^{a}t 425,
           F-91405 Orsay Cedex, France.}
\email{fournais@imf.au.dk}
%    \thanks will become a 1st page footnote.
\thanks{S. Fournais acknowledges support from the
         Carlsberg Foundation.} 

%    Information for second author
\author[M. Hoffmann-Ostenhof]{M. Hoffmann-Ostenhof}
\address{M. Hoffmann-Ostenhof: Institut f\"{u}r Mathematik,
         Strudlhofgasse 4,
         Universit\"at Wien, A-1090 Vienna, Austria.}
\email{mhoffman@esi.ac.at}
%\thanks{}

\author[T. Hoffmann-Ostenhof]{T. Hoffmann-Ostenhof}
\address{T. Hoffmann-Ostenhof: Institut f\"ur Theoretische
          Chemie,
           W\"ahringer\-strasse 17,
           Universit\"at Wien,
           A-1090 Vienna,
           Austria.}
\email{thoffman@esi.ac.at}

\author[T. \O stergaard
S\o rensen]{T. \O stergaard S\o rensen}
\address{T. \O stergaard S\o rensen: Department of Mathematical Sciences,
           Aalborg University,
           Fredrik Bajers Vej 7G,
           DK-9220 Aalborg East, Denmark.}
\email{sorensen@math.auc.dk}

%    General info
\date{March 8, 2002}
\copyrightinfo{2002}{by the authors. This article may be reproduced in
  its entirety for non-commercial purposes.} 

\begin{abstract} We prove that the electronic density of
atomic and molecular eigenfunctions is smooth away from the
nuclei. The result is proved 
without decay assumptions on the eigenfunctions.
\end{abstract}

\maketitle

\markboth{S. FOURNAIS, M. AND T. HOFFMANN-OSTENHOF, AND T. \O
STERGAARD S\O RENSEN}{ON REGULARITY OF THE DENSITY OF ATOMIC
EIGENFUNCTIONS}

\section{Introduction and main results}

We consider an $N$-electron molecule with $L$ fixed nuclei whose
non-relativistic Hamiltonian is given by
\begin{multline}
   \label{eq:Hmol}
   H=H_{N,L}(\mathbf R,\mathbf
Z)=\sum_{j=1}^N\left({}-\Delta_j-\sum_{l=1}^L
   \frac{Z_l}{|x_j-R_l|}\right)\\
   +\sum_{1\le i<j\le N}\frac{1}{|x_i-x_j|}+ \sum_{1\le l<k\le L}\frac{Z_lZ_k}
   {|R_l-R_k|},
\end{multline}
where $\mathbf R=(R_1,R_2,\dots ,R_L)\in\mathbb R^{3L}$, \(R_{l}\neq R_{k}\)
for \(k\neq l\),
denote the positions of the $L$ nuclei
whose positive charges are given by $\mathbf Z=(Z_1, Z_2,\dots,Z_L)$.
The positions of the $N$ electrons are denoted by
$(x_1,x_2,\dots,x_N)\in\mathbb R^{3N}$, where $x_j$ denotes
the position of the $j$-th electron in $\mathbb R^3$ and
$\Delta=\sum_{j=1}^N\Delta_j$ is  the $3N$-dimensional
Laplacian.

For shortness, we will sometimes write \eqref{eq:Hmol} as
$$
H = -\Delta + V.
$$
The operator
$H$ is selfadjoint on $L^2({\mathbb R}^{3N})$ with operator
domain $W^{2,2}(\mathbb R^{3N})$. We consider the eigenvalue
problem
\begin{align}
\label{eq:eigen}
H \psi = E \psi,
\end{align}
with $\psi\in L^2({\mathbb R}^{3N})$ and $E$ the real eigenvalue.
We will assume, without loss, 
%of generality, 
that the eigenfunction $\psi$ is real valued.
%with $\psi$ a real valued eigenfunction in $L^2({\mathbb
%R}^{3N})$ and $E$ the real eigenvalue. 
For $x \in {\mathbb R}^{3}$ define
$$
\psi_j(x) =  \psi(x_1,\ldots,x_{j-1},x,x_{j+1},\ldots,x_{N}),
$$
and
$$
d\hat{x}_j = dx_1\cdots dx_{j-1} dx_{j+1}\cdots dx_{N}.
$$
Then the density
associated with $\psi$ is defined as:
\begin{equation}
\label{eq:rho}
   \rho(x)=\sum_{j=1}^{N}\int \psi_j^2\,d\hat x_j.
\end{equation}
\begin{theorem}
\label{thm:density_smooth}
Let \(\psi\) satisfy \eqref{eq:eigen}. 
Then the associated density \(\rho\) 
satisfies
\begin{align}
  \rho&\in C^{\infty}({\mathbb R}^{3}\setminus
\{R_1,\ldots,R_L\}).
\end{align}
\end{theorem}
\begin{remark}
We remark that analogous natural smoothness results hold for the
$2$-electron density and the $1$-electron density matrix, 
see~\cite{FHOS1}.
\end{remark}
\begin{remark}
The electron density plays the central role in various approximation
schemes for the full $3N$-dimensional Schr\"{o}dinger equation for
$N$-electron atoms and molecules (Thomas-Fermi, Hartree-Fock, and
Density Functional Theory).

Strangely enough, the density $\rho$ has only rarely been the subject
of mathematical analysis, see~\cite{FHOS1}, \cite{HOS1}, and
references therein. Our present results are not surprising---surprising
is the fact that it has not been shown some decades ago. Since the
eigenfunction $\psi$ is real analytic away from the singularities of
the potential one would even expect $\rho$ also to be real analytic
away from the nuclei.
\end{remark}
\begin{remark}
Theorem~\ref{thm:density_smooth} is an improvement of
\cite[Theorem 1]{FHOS1} since we have no decay
assumptions on
$\psi$. This improvement will come about by using Sobolev
spaces instead of H\"{o}lder spaces.
\end{remark}

For simplicity we will here only explicitly consider the case
of atoms, i.e. we take $L=1$ and can without loss
of generality define $R_1 =0$.
To prove Theorem~\ref{thm:density_smooth} it clearly suffices to prove
\begin{align}
\label{eq:newRho}
\rho\in C^{\infty}(\mathbb R^{3}\setminus\{0\})
\end{align}
with $\rho$ redefined by \(\rho(x)=\int
\psi^2(x,x_{2},\ldots,x_{N})\,d x_2 \cdots dx_N\),
which will be done below.

Let us recall the definition of
Sobolev spaces (with integer exponent).
\begin{definition}
For $k \in {\mathbb N}$ and an open set $\Omega \subset
{\mathbb R}^n$ the Sobolev space
$W^{k,2}(\Omega)$ is defined as
$$
W^{k,2}(\Omega) = \{ f \in L^2(\Omega) \big{|} \,
\partial^{\alpha} f \in L^2(\Omega) \text{ for all } \alpha
\in {\mathbb N}^n, |\alpha| \leq k \}.
$$
\end{definition}

\section{Differentiation of \(\psi\) `along' singularities of \(V\)}

The key to the smoothness of the density is the
understanding that one is allowed to differentiate `along' 
singularities of the potential. The precise statement of
this is given in Lemma ~\ref{lem:diff_parallel}. The
statement and proof of this lemma is analogous to a similar
result in \cite[Proposition 2]{FHOS1}, the difference
being that we here measure regularity of functions by
demanding that they be in $W^{2,2}$, whereas the analogous
space in \cite[Proposition 2]{FHOS1} was the
Lipschitz continuous functions $C^{0,1}$.

\begin{definition}
Let $\Omega \subset {\mathbb R}^n$ be an open set. We define
${\mathcal B}^{\infty}(\Omega)$ as follows
$$
{\mathcal B}^{\infty}(\Omega) = \{ f\in C^{\infty}(\mathbb R^{n})
\big{|}  \,\supp f \subset
\Omega \text{ and }
\partial^{\alpha} f \in L^{\infty}(\mathbb R^{n}) \text{ for all }
\alpha \in {\mathbb N}^n \}.
$$
\end{definition}

\begin{lemma}
\label{lem:diff_parallel}
Let $P,Q$ be a partition of $\{1,\ldots,N\}$ satisfying
$$
P \neq \emptyset, \,\,\,\,\,\,\,\,\,\,\,\,
P\cap Q = \emptyset \,\,\,\,\,\,\,\,\,\,\,\,
P \cup Q =\{1,\ldots,N\}
$$
Define, for $P,Q$ as above and $\epsilon > 0$
\begin{align}
   \label{eq:def_U_P}
   U_P(\epsilon) = \Big\{ (x_1,\ldots,x_N) \in {\mathbb
R}^{3N}\,\Big|&\, |x_j| >
  \epsilon \mbox{ for } j \in P, \nonumber \\
  & |x_j-x_k| > \epsilon \mbox{ for }
  j \in P, k \in Q \Big\}.
\end{align}
Define also
$$
x_P = \frac{1}{\sqrt{|P|}} \sum_{j\in P} x_j,
$$
and let $T$ be any orthogonal transformation of ${\mathbb
R}^{3N}$ such that $T(x_1,\ldots,x_N) = (x_P,x')$ with
$x'\in {\mathbb R}^{3N-3}$. Let 
$\psi$ satisfy \eqref{eq:eigen} and
let $\phi \in {\mathcal B}^{\infty}(U_P(\epsilon))$. Then
$$
\partial^{\alpha}_{x_P} ((\phi\psi)\circ T^*) \in
W^{2,2}({\mathbb R}^{3N})
\text{ for all }
\alpha \in {\mathbb N}^3.
$$
\end{lemma}

We may assume without loss of generality that $P =
\{1,\ldots,N_1\}$, with $N_1 \leq N$. Then the orthogonal
transformation $T$ can be written as:
\begin{align}
\label{eq:T}
   T=\left(
       \begin{array}{cccccc}
         \frac{1}{\sqrt{N_1}} & \cdots& \frac{1}{\sqrt{N_1}}& 0 &
         \cdots & 0 \\
        t_{1} & \cdots& t_{N_{1}}&\cdots&\cdots & t_{N}
      \end{array}
     \right),
\end{align}
with the first row being understood as $3\times 3$ matrices---first
$N_1$ repetitions of $\frac{1}{\sqrt{N_1}}I_3$ and then $N-N_1$
repetitions of the $3 \times 3$ $0$-matrix. The remaining part of the
matrix,
\begin{align}
\label{eq:T_rows}
   \tilde{T} = \left(
    \begin{array}{ccc}
       t_1&\cdots&t_N \\
     \end{array} \right) \text{ with } t_j \in M_{3N-3,3}({\mathbb R}), 
\end{align}
is such that the complete matrix $T$ is orthogonal.

Let us write $\tilde{\psi} = \psi \circ T^*$. Notice that,
since $\Delta$ is invariant under orthogonal
transformations, $\tilde{\psi}$ satisfies
\begin{equation}
\label{eq:tilde}
-\Delta \tilde{\psi} + (V\circ T^*) \tilde{\psi} = E
\tilde{\psi}.
\end{equation}
We can now prove Lemma~\ref{lem:diff_parallel}.

\begin{proof}
We will prove that for all $\phi \in
{\mathcal
B}^{\infty}(U_P(\epsilon(1-2^{-k})))$ we
have
$\partial^{\gamma}_{x_P} (\tilde\phi \tilde{\psi} )
\in  W^{2,2}({\mathbb R}^{3N})$
for all $\gamma \in {\mathbb N}^3$ with $|\gamma| \leq k$, with
\(\tilde\phi=\phi\circ T^*\).
The proof is by induction with respect to
$k$. For $k=0$ there is nothing to prove since
${\mathcal D}(H) = W^{2,2}({\mathbb R}^{3N})$.
Let $k \geq 0$ and suppose that $\partial^{\gamma}_{x_P}
(\tilde\phi \tilde{\psi} )
\in  W^{2,2}({\mathbb R}^{3N})$ for all
$\phi \in {\mathcal
B}^{\infty}(U_P(\epsilon(1-2^{-k})))$ and all
$\gamma \in {\mathbb N}^3$ with $|\gamma|\leq k$.
Take now
$\gamma \in {\mathbb N}^3$
with $|\gamma|= k+1$. Let $\phi
\in {\mathcal
B}^{\infty}(U_P(\epsilon(1-2^{-(k+1)})))$. From the
eigenfunction equation \eqref{eq:tilde} for
$\tilde{\psi}$ we get
\begin{align}
\label{eq:Schroedinger_loc}
-\Delta (\tilde\phi \tilde{\psi})
&= (-2 \nabla \tilde\phi \nabla \tilde{\psi} - \tilde{\psi} \Delta \tilde\phi + E \tilde\phi
\tilde{\psi}) \nonumber \\
& - \sum_{j=1}^N \left( \frac{-Z}{|x_j|} \circ
T^*\right) (\tilde\phi
\tilde{\psi}) -
\sum_{1\leq j<k\leq N} \left( \frac{1}{|x_j-x_k|}\circ
T^* \right) (\tilde\phi
\tilde{\psi}).
\end{align}
Now we differentiate equation \eqref{eq:Schroedinger_loc} with respect
to \(x_{P}\) (in the distributional sense). 

By the induction hypothesis it is clear that
$$
\partial^{\gamma}_{x_P} (-2 \nabla \tilde\phi \nabla \tilde{\psi} -
\tilde{\psi}
\Delta \tilde\phi+ E \tilde\phi \tilde{\psi}) \in L^2({\mathbb R}^{3N}).
$$
Notice that $\partial_{x_P}$ commutes with
$|x_j|^{-1}\circ T^*$ for $j \in Q$ and with
$|x_j-x_k|^{-1}\circ T^*$ for $j,k \in P$ or $j,k \in Q$.
Furthermore, on $\supp \tilde\phi$ the remaining parts of the
potential, $|x_j|^{-1}\circ T^*$ for $j \in P$ and
$|x_j-x_k|^{-1}\circ T^*$ with $j\in P$, $k \in Q$, are
bounded functions with bounded derivatives of arbitrary
order, so we get, using the induction
hypothesis that
$$
\partial_{x_P}^{\gamma} ((V\circ T^*) \tilde\phi \tilde{\psi}) =
    (V\circ T^*) \partial_{x_P}^{\gamma}(\tilde\phi \tilde{\psi}) +
g,
$$
where $g \in W^{2,2}({\mathbb R}^{3N})$.
Furthermore, by assumption we have
$\partial_{x_P}^{\gamma}(\tilde\phi \tilde{\psi}) \in W^{1,2}({\mathbb
R}^{3N})$. Therefore we obtain 
$(V\circ T^*) (\partial_{x_P}^{\gamma}(\tilde\phi \tilde{\psi}))
\in L^2({\mathbb R}^{3N})$ since the operators \(|x_{j}|^{-1}\) and
  \(|x_{j}-x_{k}|^{-1}\) are bounded from
  \(W^{1,2}(\mathbb R^{3N})\) to \(L^{2}(\mathbb R^{3N})\) (see for
  instance~\cite[p.\ 169]{ReSi}).

Thus, in the sense of distributions,
~\eqref{eq:Schroedinger_loc} implies
$$
  \Delta \partial_{x_P}^{\gamma}(\tilde\phi \tilde{\psi}) \in
L^2({\mathbb R}^{3N}).
$$
Via standard elliptic regularity results,
this implies that $\partial_{x_P}^{\gamma}(\tilde\phi
\tilde{\psi})
\in W^{2,2}({\mathbb R}^{3N})$ and finishes
the proof.
\end{proof}

\section{The differentiation of \(\rho\)}

In this section we perform the differentiation of the
density, as stated in ~\eqref{eq:newRho}.
Let $\phi \in C_0^{\infty}({\mathbb R}^{3} \setminus
\{0\})$. We will prove that
\begin{align}
\label{eq:L-1}
\partial_x^{\gamma} (\phi^2 \rho) \in L^1({\mathbb R}^3)
\text{ for all } \gamma \in {\mathbb N}^3.
\end{align}
By a Sobolev embedding theorem, we therefore get $\phi^2 \rho
\in C^{\infty}({\mathbb R}^3)$, which implies \eqref{eq:newRho},
since
$\phi \in C_0^{\infty}({\mathbb R}^{3} \setminus
\{0\})$ was arbitrary.

We use the partition of unity introduced in
\cite{FHOS1}. Let $R>0$ be such that $\{x \in {\mathbb
R}^{3} \big{|}\, |x| \leq R \}  \cap \supp
\phi = \emptyset$.
Let $\chi_1,\chi_2$ be a partition of unity in ${\mathbb R}_{+}$:
$\chi_1+\chi_2 =1$, with $\chi_1(x) = 1 $ on $[0,R/(4N)]$,
$\supp \chi_1
\subset [0,R/(2N)]$ and
$\chi_j \in C^{\infty}({\mathbb R_{+}})$ for $j=1,2$.
We combine the $\chi_j$'s to make a partition of unity in
${\mathbb R}^{3N}$. Obviously,
\begin{align*}
   1 =
   \prod_{1\leq j<k\leq N}
\Big( \chi_1\big(|x_j-x_k|\big)+\chi_2\big(|x_j-x_k|\big) \Big).
\end{align*}
Multiplying out the above product, we get sums of products of
$\chi_1$'s and $\chi_2$'s. We introduce the following index
sets to  control these sums:
Define first
\begin{align*}
   M=\{ (j,k) \in \{ 1,\ldots,N\}^2\,|\, j <k \},
\end{align*}
and let
\begin{eqnarray*}
    I \subset M, \,\,\,\,\,\,\,
    J  = M\setminus I.
\end{eqnarray*}
Now define, for each pair $I,J$ as above,
\begin{align*}
    \phi_{I}^2({\mathbf x}) &= \Big(
    \prod_{(j,k)\in I} \chi_1\big(|x_j-x_k|\big) \Big)
\Big(\prod_{(j,k)\in J} \chi_2\big(|x_j-x_k|\big) \Big).
\end{align*}
Then we get
\begin{align*}
   1 = \prod_{1\leq j<k\leq N}\big(\chi_1+\chi_2\big)\big(|x_j-x_k|\big) =
   \sum_{I\subset M} \phi_{I}^2({\mathbf x}),
\end{align*}
where the sum is over all subsets $ I \subset M$.

Therefore we have, with $g_I = \psi^2 \phi_I^2$,
\begin{align*}
   \rho(x_1)
   = \sum_{I\subset M} \int g_I
    (x_1,x_2,\ldots,x_N)\,dx_2 \cdots\,dx_N
                \equiv \sum_{I\subset M} \rho_I(x_1).
\end{align*}

To verify \eqref{eq:L-1} we have to prove that for every \(I\subset M\),
$
\partial_x^{\gamma} (\phi^2 \rho_I) \in L^1({\mathbb R}^3)$
for
  all $\gamma \in {\mathbb N}^3$. As in \cite{FHOS1}, for
each $I \subset M$ we can choose a $P$ (depending on $I$),
with $1\in P$, and such that
$\supp
\phi(x_1)
\phi_I(x_1,\ldots,x_N) \subset U_P(R/4N)$ (using that
$\supp\phi\cap\{\,|x|\leq R\}=\emptyset$).
This will be essential in the considerations below in order to apply
Lemma~\ref{lem:diff_parallel}.

Let $u \in C_0^{\infty}({\mathbb R}^3)$. Choose $f \in
C_0^{\infty}({\mathbb R}^{3N-3})$, with $f\equiv 1$ for $|x|
< 1$, and let $f_K(x) = f(x/K)$, \(K>0\). We calculate
\begin{eqnarray}
\label{eq:distribution}
\int_{{\mathbb R}^{3}} (\partial_{x_1}^{\gamma} u)(x_1)
(\phi^2
\rho_I)(x_1) \, dx_1 &=&
\int_{{\mathbb R}^{3N}} (\partial_{x_1}^{\gamma} u)(x_1)
(\phi^2 g_I)(x_1,\ldots,x_N)\,dx_1\cdots dx_N \nonumber \\
&=&
N_{1}^{-|\gamma|/2}
\int_{{\mathbb R}^{3N}} (\partial_{x_P}^{\gamma}
\tilde{u}) (\tilde{\phi}^2 \tilde{g}_I)(x_P,x')\,dx_P dx',
\end{eqnarray}
according to the coordinate transform given in
Lemma~\ref{lem:diff_parallel} as specified in~\eqref{eq:T}
and~\eqref{eq:T_rows}. Here
$\tilde{u} = u
\circ T^*$ and similarly for the other functions. By Lebesgue
integration theory it follows that~\eqref{eq:distribution}
equals
\begin{align}
\label{eq:intro_hR}
N_{1}^{-|\gamma|/2}
\lim_{K\rightarrow \infty}
\int_{{\mathbb
R}^{3N}} (h_K \tilde{\phi}^2 \tilde{g}_I)(x_P,x')\,dx_P dx',
\end{align}
with $h_K(x_P,x') = (\partial_{x_P}^{\gamma}
\tilde{u})(x_P,x') f_K(x')$.\\
Noting that $x_P = \sqrt{N_1}(x_1-t_1^* x')$, \(t_1^*\) being the
transposed of \(t_1\), and that $u$
and $f_K$ are $C_0^{\infty}$ functions, $h_K \in
C_0^{\infty}({\mathbb R}^{3N})$. Hence ~\eqref{eq:intro_hR} equals 
\begin{align}
\label{eq:diff_dist}
(-1)^{|\gamma|}N_{1}^{-|\gamma|/2}
\lim_{K\rightarrow \infty}
\int_{{\mathbb
R}^{3N}} \tilde{u} f_K v\,dx_P dx',
\end{align}
with $v(x_P,x') = \partial_{x_P}^{\gamma} (\tilde{\phi}^2
\tilde{g}_I)(x_P,x')$. Here the derivative is taken in the
distributional sense.

Suppose we have shown that $v\in L^1({\mathbb R}^{3N})$, then
by Lebesgue integration theory ~\eqref{eq:diff_dist} equals
\begin{align}
\label{eq:removing_limit}
(-1)^{|\gamma|}N_{1}^{-|\gamma|/2}
\int_{{\mathbb
R}^{3N}} \tilde{u}  v\,dx_P dx',
\end{align}
and by transforming again and applying Fubini's
theorem ~\eqref{eq:removing_limit} equals
\begin{align}
\label{eq:integrating_some_out}
(-1)^{|\gamma|}N_{1}^{-|\gamma|/2}
\int_{{\mathbb
R}^{3}} \tilde{u}  w\,dx_1,
\end{align}
with $w(x_1) = \int_{{\mathbb R}^{3N-3}} (v\circ
T)(x_1,\ldots, x_N)\,dx_2 \cdots dx_N$, \(w\in L^{1}({\mathbb R}^{3})\).
Combining
~\eqref{eq:distribution} --
~\eqref{eq:integrating_some_out}, 
\eqref{eq:L-1} follows.

So to finish the proof of \eqref{eq:newRho}
it remains to verify that $v \in L^1({\mathbb R}^{3N})$,
which will follow via Lemma~\ref{lem:diff_parallel}. In
order to apply the lemma we need the following.

\begin{prop}
\label{prop:Leibniz}
Let $u,v \in L^2({\mathbb R}^{n})$ and
$\partial_{x_1} u, \partial_{x_1} v \in L^2({\mathbb
R}^{n})$ (distributional sense). Then
$$
\partial_{x_1} (uv) = (\partial_{x_1} u) v + u\partial_{x_1}
v, \text{ in the distributional sense}.
$$
\end{prop}

Proposition~\ref{prop:Leibniz} is easy to prove by standard
density arguments.

Remember that
$v=\partial_{x_P}^{\gamma}(\tilde\phi^2
\tilde{\phi}_I^2 \tilde{\psi}^2)$, and that $P$ was chosen
such that $\supp \phi \phi_I \subset U_P(\epsilon)$ for 
\(\epsilon=R/4N\). Let now $v_1 = 
\phi \phi_I \psi$. Then $v = \partial_{x_{P}}^{\gamma}(\tilde v_{1}^{2})$.
Since
$\tilde{\phi} \tilde{\phi}_I \in {\mathcal
B}^{\infty}({\mathbb R}^{3N})$ 
it follows via Lemma~\ref{lem:diff_parallel} and Leibniz' rule that
$\partial_{x_P}^{\alpha}(\tilde v_1) \in L^2({\mathbb R}^{3N})$
for all 
$\alpha
\in {\mathbb N}^3$.
 From this and Proposition~\ref{prop:Leibniz} it is easy to
prove by induction that
$$
\partial_{x_P}^{\gamma}
\tilde{v}_1^2
=
\sum_{\alpha+\beta=\gamma}
(\partial_{x_P}^{\alpha} \tilde{v}_1)
(\partial_{x_P}^{\beta} \tilde{v}_1) \in L^1({\mathbb
R}^{3N}).
$$
This finishes the proof
of \eqref{eq:newRho} and hence of Theorem~\ref{thm:density_smooth} for
the atomic case.

\def\dbar{\leavevmode\hbox to 0pt{\hskip.2ex \accent"16\hss}d}

\end{document}